\documentclass[sigconf]{acmart} 

\usepackage{booktabs}
\usepackage{multirow}

\AtBeginDocument{%
  }

\copyrightyear{2024}
\acmYear{2024}
\setcopyright{rightsretained}
\acmConference[VRST '24]{30th ACM Symposium on Virtual Reality Software and Technology}{October 9--11, 2024}{Trier, Germany}
\acmBooktitle{30th ACM Symposium on Virtual Reality Software and Technology (VRST '24), October 9--11, 2024, Trier, Germany}\acmDOI{10.1145/3641825.3689510}
\acmISBN{979-8-4007-0535-9/24/10}

\begin{document}

\title{VR4UrbanDev: An Immersive Virtual Reality Experience for Energy Data Visualization}

\author{Saeed Safikhani}
\affiliation{
  \institution{Graz University of Technology}
  \city{Graz}
  \country{Austria}
  }
  \email{s.safikhani@tugraz.at}

\author{Georg Arbesser-Rastburg}
\author{Anna Schreuer}
\author{Jürgen Suschek-Berger}
\author{Hermann Edtmayer}
\affiliation{
  \institution{Graz University of Technology}
  \city{Graz}
  \country{Austria}
  }

\author{Johanna Pirker}
\affiliation{
  \institution{Graz University of Technology}
  \city{Graz}
  \country{Austria}}
\affiliation{
  \institution{Ludwig-Maximilians-Universität München}
  \city{Munich}
  \country{Germany}
  }

\renewcommand{\shortauthors}{Safikhani et al.}

\newcommand{\cs}{\textbf{CS} }
\newcommand{\civil}{\textbf{Civil} }

\begin{abstract}

In this demonstration paper, we present our interactive virtual reality (VR) experience, which has been designed to facilitate interaction with energy-related information. This experience consists of two main modes: the world in miniature for large-scale and first-person for real-world scale visualizations. Additionally, we presented our approach to potential target groups in interviews. The results of these interviews can help developers for future implementation considering the requirements of each group.
\end{abstract}

\keywords{Virtual Reality, Building Information Modeling, Data Visualization}

\maketitle

\section{Introduction}
\noindent The rise in popularity of consumer-level head-mounted displays (HMD) is attracting interest in immersive virtual reality (VR) applications from a variety of entertainment and research fields. Building Information Modeling (BIM) holds significant promise in architecture, engineering, and construction (AEC) \cite{safikhani2022immersive}. 
Previous studies showed that VR can be beneficial for education/training \cite{Peterson2021, ahmed2020integrating}, design/data exchange \cite{calderon2019comparing}, and management/collaboration \cite{nasrazadani2020implementation}. The current trend in engineering is moving towards considering energy efficiency in every design. While VR has been utilized in various applications in AEC, we have not yet seen a comprehensive implementation of VR for energy data visualization and interactions. In this work, we present a VR application that allows for demonstrating different energy-related information, using either real-time or historical data. We used data gathered from the university campus as proof of concept for our application. However, this application can be easily expanded to a much larger area. Our campus has integrated various IoT systems to collect energy-related data, including electrical energy and energy used for heating and cooling. Leveraging this data, we developed an initial prototype for VR-enabled data visualization and data-driven planning. We then interviewed different user groups to gather preliminary feedback and identify areas for improvement and further application of our approach.

\section{Technical Description}
\noindent We utilized Unreal Engine (UE) 5.3 for creating our VR interaction logic, visuals, and data transfer. UE provides us with realistic rendering features and a comprehensive set of tools, enabling us to develop different interactions in a pleasant visual interface. As the project required us to include large-scale map information, we integrated the Cesium plugin for UE. This plugin allows for integrating real-world maps based on sources such as Microsoft Bing and Google Maps. 
We utilized the OpenXR plugin for seamless VR communication between the HMD and UE to extend the range of supported devices. As a result, the current version of the project supports devices such as SteamVR, Meta Quest, and Windows Mixed Reality.

\section{Virtual Environment}

\begin{figure}[h]
     \centering
     \includegraphics[width = 8 cm]{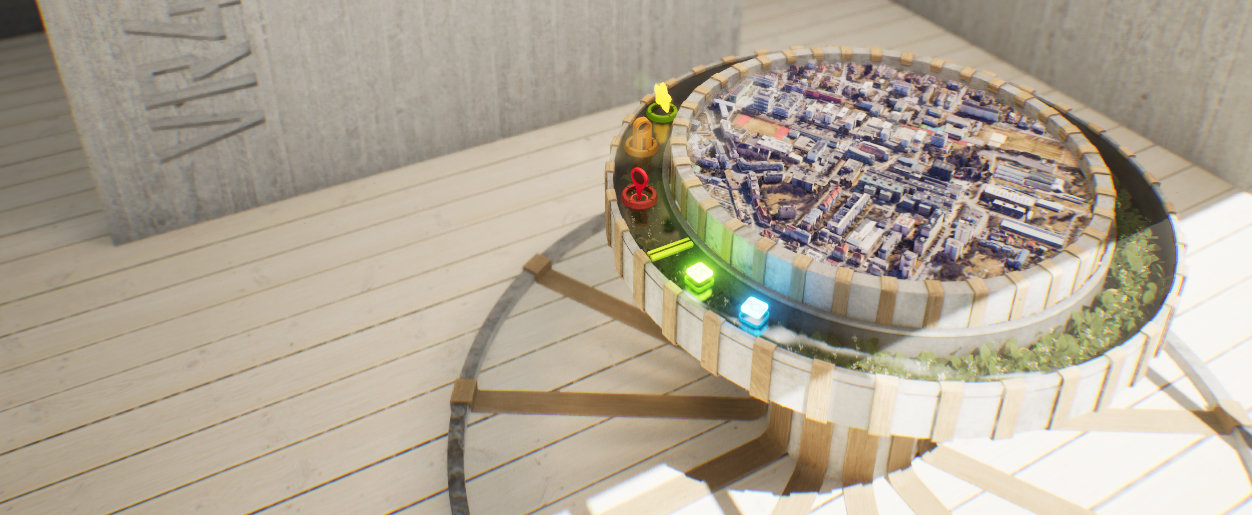}
     \includegraphics[width = 8 cm]{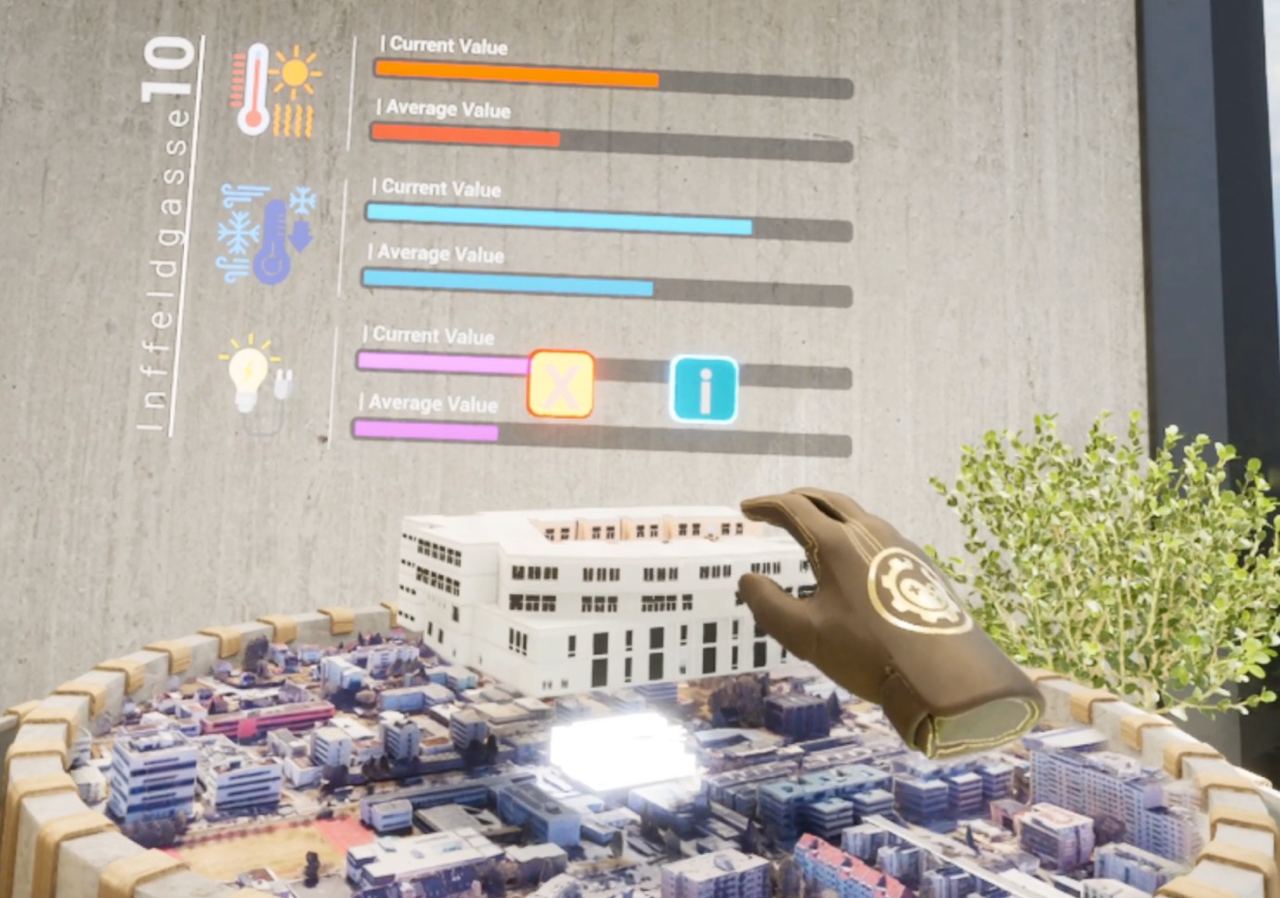}
     \Description{Top figure depicts a room featuring a round, two-tiered table. The larger, lower level of the table is equipped with interactive buttons, while the upper level showcases a 3D map of a city. In the Bottom figure, a VR user interacts with a building on the 3D city map displayed on the round table. Simultaneously, the results of this interaction are projected onto the wall behind the table. These results display various graphs illustrating energy consumption for cooling, heating, and electricity.}
     \caption{(Top) The environment of the world in miniature mode with interaction desk, (Bottom) Visualizing real-time energy data in the world in miniature mode}
     \label{fig:room_overview}
\end{figure}

\noindent Our VR application includes two main modes: 

\noindent\textbf{The world in miniature mode:} is set in a minimalist room and is designed to provide a large-scale visualization and interaction with the map (Figure \ref{fig:room_overview}-top). 
We used a circular desk in this room as the main interaction area.
In the center of the room, there is a circular desk called the "interaction desk" that serves as the main interaction point for the map.
Users can scale and pan the map using this desk. 
\noindent The interaction desk is divided into two layers: the top layer for the map and the bottom layer for gadgets and visualization mode selection. 
Users can rotate these layers separately by grabbing and moving the edge of each layer.

\noindent The buildings on the campus map can be interacted with by hovering over them and touching them. Doing so will open up a detailed model of the building, with the option to access additional information by pressing the information button. Depending on the current visualization mode and gadget, certain information will be displayed on the walls of the room (Figure \ref{fig:room_overview}-bottom).

\noindent The bottom layer consists of two areas: gadgets and visualization modes (Figure \ref{fig:room_overview}-top). The gadgets area is a placeholder for different gadgets that can be used for various interactions with the map. For example, users can place a measuring gadget on the map to measure the distance between two points. Another available gadget is world-teleportation. By placing this gadget on the map, users can be transferred to that location in first-person mode.
The visualization mode area contains several interactive buttons for switching or adding visualizations to the map. For instance, users can activate the isolation mode to highlight the campus area or use the point of interest mode to display important locations on the map.

\noindent\textbf{The first-person mode:} is specifically designed to offer users a real-world scale experience. To access this mode, users need to use a dedicated gadget in the miniature mode. Once in the mode, users can view different energy-related information for each building in a specific area. As moving around in real-world scale can be time-consuming, we will provide users with a local map option to help them teleport quickly to a specific location on the map.

\section{User Interface and Interactions}
\noindent Through our previous experiences in developing VR interactions, we learned that users prefer integrated interactions over a copy of traditional desktop interactions in VR. However, using too many 3D interactions with different functionalities can be challenging for new users to learn and cumbersome to use in the long term. To address this, we decided to simplify the design of interactions and make them consistent throughout the entire experience.
Our main interaction is grabbing objects and manipulating them accordingly, along with pressing physically-behaved buttons. The grab interaction is generalized throughout the entire experience and can be performed with one or two hands for different applications. Additionally, physical buttons behave similarly to real-world buttons, allowing users to immediately understand what to expect from them.
We considered two different types of locomotion in addition to physical movement, namely teleportation and smooth movement. The smooth movement system is similar to conventional game locomotion using joysticks which provides a continuous experience of movement but might result in motion sickness. On the other hand, teleportation has less chance of causing motion sickness, but it might lead to disorientation. Users can freely choose their preferred mode based on their sensitivity to motion sickness in both modes.
This project will be expanded with additional features such as different data visualization modes and building manipulation.

\section{Interview With Experts}
To evaluate our approach, we conducted interviews with stakeholder groups that may use the environment for their specific tasks:\\
\noindent \textbf{Energy researchers:} interactive data visualization for research\\
\noindent \textbf{Planners:} visualizing design decisions and external collaboration\\
\noindent \textbf{Facility managers:} monitoring hidden infrastructures \\
We carried out a total of 10 interviews, each structured into three parts. The first part covered the interviewee's job, their use of software tools for visualizing energy data, and prior VR experience. The second part involved testing our VR experience. Finally, participants provided feedback on the experience and its usability, discussing potential integration into their workflow.

\noindent Initial results indicate that participants found the tool user-friendly and appreciated the virtual environment, though some struggled to see how it could provide added value to their daily work practices. Potential applications ranged from high-level planning, such as outdoor area design, to detailed analysis and fault detection, like visualizing indoor sensor positions and comparing various data types to ensure system functionality. We will incorporate the feedback and ideas based on the analysis of this interview into our project.
This will include simplified and streamlined visualizations, improvements to the interaction and locomotion systems, as well as visualizing the influence of design decisions.
In the future, we plan to conduct another round of interviews with more participants to further evaluate our approach.

\noindent This work was supported by the Austrian Research Promotion Agency (FFG) program \textit{City of Tomorrow}, project no. FO999893555.

\bibliographystyle{ACM-Reference-Format}
\bibliography{01_bib}

\end{document}